\documentclass[%
 reprint,
 amsmath,amssymb,
 aps, pra
]{revtex4-2}

\usepackage[utf8]{inputenc}
\usepackage{amsmath}% For equations
\usepackage{amsfonts}
\usepackage{amssymb}
\usepackage{graphicx} % Include figure files
\usepackage{dcolumn}% Align table columns on decimal point
\usepackage{bm}% Bold math
\usepackage{braket}
\usepackage{xcolor}
\usepackage{soul}
\usepackage{float}
\definecolor{highlightcolor1}{RGB}{255,240,150} % Yellow color
\definecolor{highlightcolor2}{RGB}{173,216,230} % Light blue color

\newcommand{\be}{\begin{equation}}
\newcommand{\ee}{\end{equation}}

\begin{document}
\setcounter{page}{1} 
\title{Ultra-cold atoms quantum tunneling through single and double optical barriers}

\author{R. Eid}
\affiliation{Universit\'e Paris-Saclay, Institut d'Optique Graduate School, CNRS, Laboratoire Charles Fabry,  91127 Palaiseau, France}

\author{A. Hammond}
\affiliation{Universit\'e Paris-Saclay, Institut d'Optique Graduate School, CNRS, Laboratoire Charles Fabry,  91127 Palaiseau, France}

\author{L. Lavoine}
\affiliation{Universit\'e Paris-Saclay, Institut d'Optique Graduate School, CNRS, Laboratoire Charles Fabry,  91127 Palaiseau, France}

\author{T. Bourdel}
\email[Corresponding author: ]{thomas.bourdel@institutoptique.fr}
\affiliation{Universit\'e Paris-Saclay, Institut d'Optique Graduate School, CNRS, Laboratoire Charles Fabry,  91127 Palaiseau, France}

\date{\today}

\begin{abstract}
We realize textbook experiments on Bose-Einstein condensate tunnelling through thin repulsive potential barriers. In particular, we demonstrate atom tunnelling though a single optical barrier in the quantum  scattering regime where the De Broglie wavelength of the atoms is larger than the barrier width. Such a beam splitter can be used for atom interferometry and we study the case of two barriers creating an atomic Fabry-P\'erot cavity. Technically, the velocity of the atoms is reduced thanks to the use of a $^{39}$K Bose-Einstein condensate with no interactions. The potential barriers are created optically and their width is tunable thanks to the use of a digital micro-mirror device. In addition, our scattering experiments enable in-situ characterization of the optical aberrations of the barrier optical system.  
\end{abstract}
\maketitle

Particle quantum tunneling is a phenomenon in which a particle passes through a potential energy barrier, that according to classical mechanics should not be passable due to insufficient energy. It is a direct consequence of the wave-nature of matter and is described by the Sch\"odringer equation \cite{griffiths1995introduction}. Experimentally, it was first observed in 1957 by L. Esaki for electrons in semiconductors \cite{Esaki1973}, who then used this effect to build electronic diodes \cite{esaki1958}. The scanning tunnelling microscope is an important application based on quantum tunnelling \cite{binnig1982tunneling}. 

Equivalently, atoms can also exhibit quantum (wave) behavior, although they need to be cooled to ultralow temperatures. Ultracold atoms are indeed used both for precision measurements using matter-wave interferometers \cite{Cronin09} or for the study of quantum many-body physics \cite{Bloch:08}. Tunnelling of atoms between sites of an optical lattice is a commonly observed phenomenon \cite{Greiner2002}. In a wave\-guide configuration, beam splitting was realized through Bragg scattering in lattices \cite{Fabre11}. Quantum reflection from the attractive potential close to a solid surface was also observed \cite{Pasquini04, Shimizu01}. 

In the simple textbook experiment of an atom crossing a single potential barrier, the interesting regime of coherent splitting requires a barrier size $\sigma$ comparable to the atom De Broglie wavelength $\lambda_\textrm{dB}=h/mv$, where $m$ is the atom mass, $v$ the atom velocity, and $h$ the Planck constant. Since the minimum optical barrier size is limited by diffraction to the optical wavelength, it thus requires to reduce and control $v$ to sub-millimeter per second velocities. As a consequence, there have been only a few experiments on quantum tunnelling through optical barriers with Bose-Einstein condensate (BEC) \cite{Pasquini_2006,Marchant16,Billy_2007, Ramos_2018, ramos2019measuring}. The low velocity thin barrier regime, characterized by $\lambda_\textrm{dB}\ll \sigma$ and also called to the quantum scattering regime, has never been clearly observed. In contrast, numerous theoretical studies have explored atomic tunneling phenomena, considering various barrier shapes and incorporating interactions within Bose-Einstein condensates (BECs) \cite{Martin_2007,Manju_2018, Lindberg23}. Matter-wave Fabry-P\'erot interference using two consecutive barriers remains to be observed with potential applications for narrowing the atomic velocity distribution in precision measurements apparatus \cite{Manju_2020}.

In the last ten years, digital micro-mirror devices (DMD) have been shown to be a great tool to impose arbitrary potentials in ultracold atom experiments \cite{Liang:09, Gaunt_2013, Gauthier:16, Tajik:19, Navon_2021}. For example, box traps permit to study gases at a constant density \cite{Gaunt_2013,Navon_2021} and donut shape traps are nice for the study of superfluid rotation \cite{Zou_2021}. Thanks to their versatility and the novel possibilities they offer, DMDs are increasingly used. The optical resolution of such setups is usually measured before their installation in the ultracold atom experiment and not characterized in-situ. 

In this paper, we perform atom tunnelling experiments through simple and double repulsive optical barriers. The barriers are generated through a DMD setup allowing the adjustment of the barrier width and position. The fine control of the atom velocity is achieved by using a $^{39}$K condensate that can be made non-interacting thanks to Feshbach tuning \cite{DErrico07}. For single barriers, we are able to distinctly show the two different regimes of scattering. When $\lambda_\textrm{dB} < \sigma$, the scattering is essentially classical. The transmission curve as a function of barrier height is close to a step function although rounded by quantum effects. When $\lambda_\textrm{dB} > \sigma$, the transmission curve corresponds to the one expected for a delta potential. It is the quantum scattering regime. We then realize double barrier potentials. Interestingly, we observe oscillations of the transmission as a function of the distance between the two barriers.  By comparison to numerical simulations, this behavior is interpreted as originating both from atomic Fabry-P\'erot interference and from optical interference due to the oscillatory behavior of the point spread-function (PSF) of the optical system. Interestingly, our results with various DMD patterns permit $in-situ$ characterization of the PSF. 

The experiments start with the production of $^{39}K$ Bose-Einstein condensates in the $|F=1, m_F=1\rangle$ by evaporation in a crossed optical dipole trap at 393\,G, where the scattering length $a$ is $\sim$130\,$a_0$, where $a_0$ is the Bohr radius \cite{Berthet20}. Subsequently, within 100\,ms, the trapping frequencies are modified to $(\omega_\perp,\omega_\parallel)/2\pi=(120,16)\,\mathrm{Hz}$ and the magnetic field is adjusted to $\sim$350\,G in close proximity to the magnetic field where we observe a collapse of the condensate. The gas is then almost non-interacting \footnote{In this regime of very low scattering length $a<0.3$\,$a_0$, the magnetic dipole interaction plays a role comparable to the contact interaction}. If the axial confinement is removed, we observe a very slow expansion of the condensate in the remaining optical waveguide. It corresponding to a mean energy of $h\times$8\,Hz or equivalently a velocity spread upon release $\Delta v_0 \sim 0.35\,$mm/s. This is only slightly higher than the kinetic energy $h \omega_\perp /8\pi$ expected from a non-interacting condensate. There is also a residual $\sim$20\% non condensed atomic fraction that expands much faster $\Delta v_\textrm{thermal} \sim 2.6\,$mm/s.

\begin{figure}[ht]
\includegraphics[width=\columnwidth]{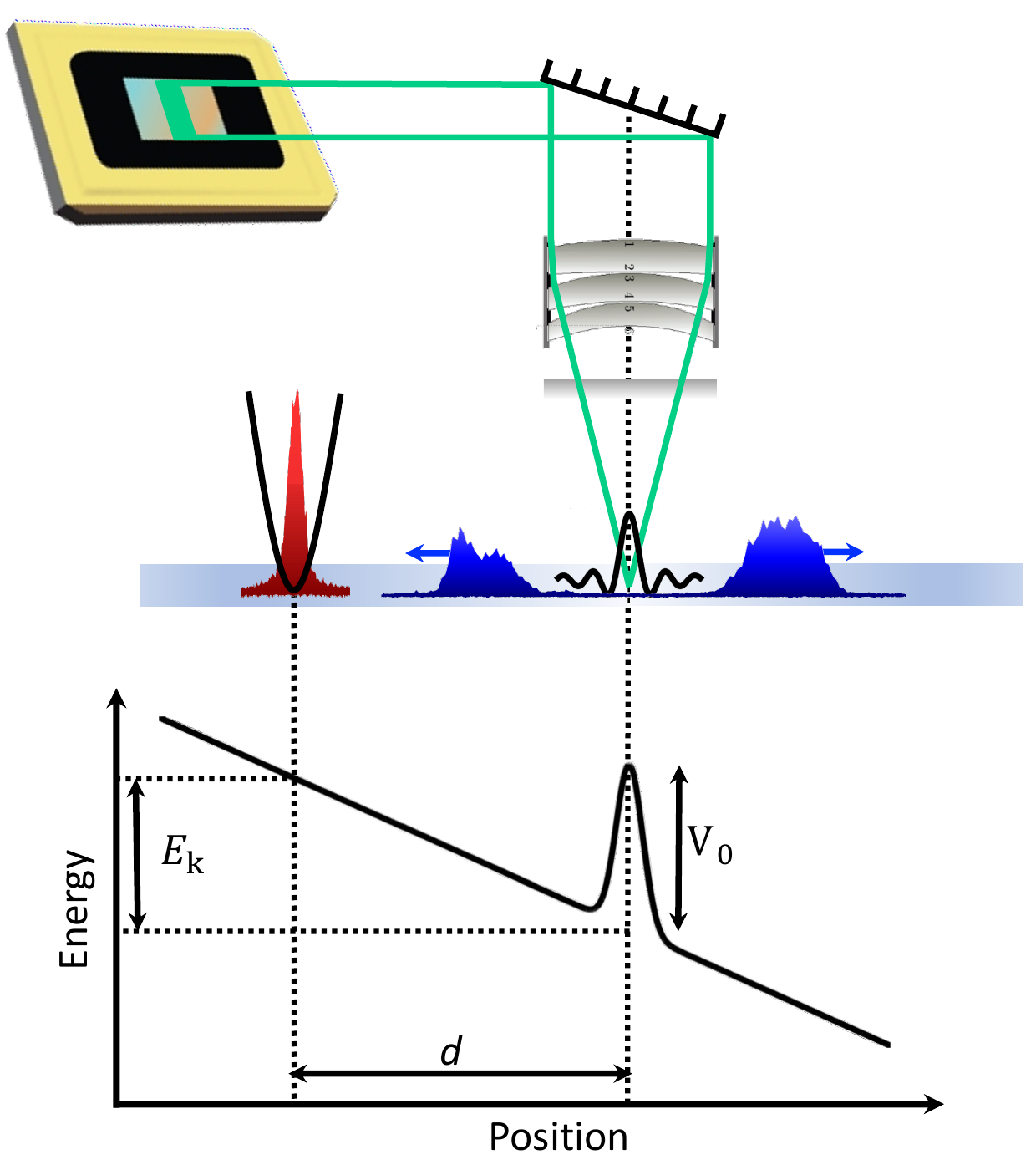}
\caption{Schematic of the experiment. A Bose-Einstein condensate is released and accelerated in an optical waveguide toward a potential barrier. This potential is made with a DMD optical setup at 532\,nm. The measured quantities are the transmitted and reflected atom numbers after the collision when the two clouds are well separated. An example of a measured longitudinal density profile is shown.   
\label{Paper_picture}}
\end{figure}

The optical barriers at 532\,nm (Verdi V18, Coherent) are created with an optical setup using a digital micro-mirror devices (DLP-6500, Texas Instrument). The DMD is composed of a matrix of 1920x1080 small square mirrors of pitch size $p=$7.56\,$\mu$m. Each of them can take two different angles $\pm 12^\circ$. The DMD is illuminated by a large collimated beam at an angle of 24$^\circ$. We then select only the most intense order of diffraction orthogonal to the DMD that corresponds to the specular reflection and image the DMD plane on the atoms through a custom-made objective. This objective is composed of 3 spherical lenses and has an overall focal length of 41.16\,mm  at 532\,nm. The numerical aperture is 0.27. The distance from the DMD to the objective is 1.5\,m such that the demagnification factor is 38.5 and the effective pixel size in the atomic plane is measured to be $p$=0.196\,$\mu$m. The objective design takes into account the glass cell and theoretically gives perfect diffraction-limited performances and an Airy function as PSF. In the following, we use one-dimensional (1D) DMD patterns made of lines perpendicular to the propagation of the atoms. In that case, a more relevant quantity is the 1D amplitude PSF which is theoretically a sinc function. Given the numerical aperture, the first zero of the sinc function should theoretically correspond to $\sigma_{\textrm{th}}=$1.0\,$\mu$m in the atomic plane (Fig.\,\ref{Paper_picture}). 

The experiment consists in releasing the Bose-Einstein condensate into an horizontal optical waveguide where the atoms are accelerated toward the optical barriers. The longitudinal acceleration is induced by a magnetic field gradient and is $g=$0.20\,m/s$^2$. The collision with a potential barrier results in partial reflection and transmission of the atomic cloud that are detected through fluorescent imaging (Fig.\,\ref{Paper_picture}). Choosing the distance $d$ from the initial trap to the barrier position, we can control the speed of the atoms when they meet the barrier. We can also vary the barrier height through an accousto-optic modulator that modifies the 532\,nm light power sent to the DMD. The barrier transmission can then be studied for different conditions. A key parameter is the ratio of the De Broglie wavelength of the atoms $\lambda_\textrm{dB}=h/mv$ to the barrier width $\sigma$. Another important parameter is the atomic velocity spread $\Delta v$ when the atom meet the barrier. Interestingly, since we keep the acceleration on for the whole sequence, $\Delta v$ is not directly given by $\Delta v_0$. The kinetic energy of an atom with an initial velocity $v_0$ at a distance $d$ from the barrier is $E_\textrm{k}=\frac{1}{2}mv^2=\frac{1}{2}mv_0^2+mgd$. For our parameters, the second term dominates such that $v \approx\sqrt{2gd}$ and $\Delta v/v\approx \Delta x/2d\ll1$ where $\Delta x \approx 3.5\,\mu$m is an estimated rms initial size of the cloud knowing the longitudinal trap frequency and the measured expansion energy.

\begin{figure}[h!]
\includegraphics[width=0.9\columnwidth]{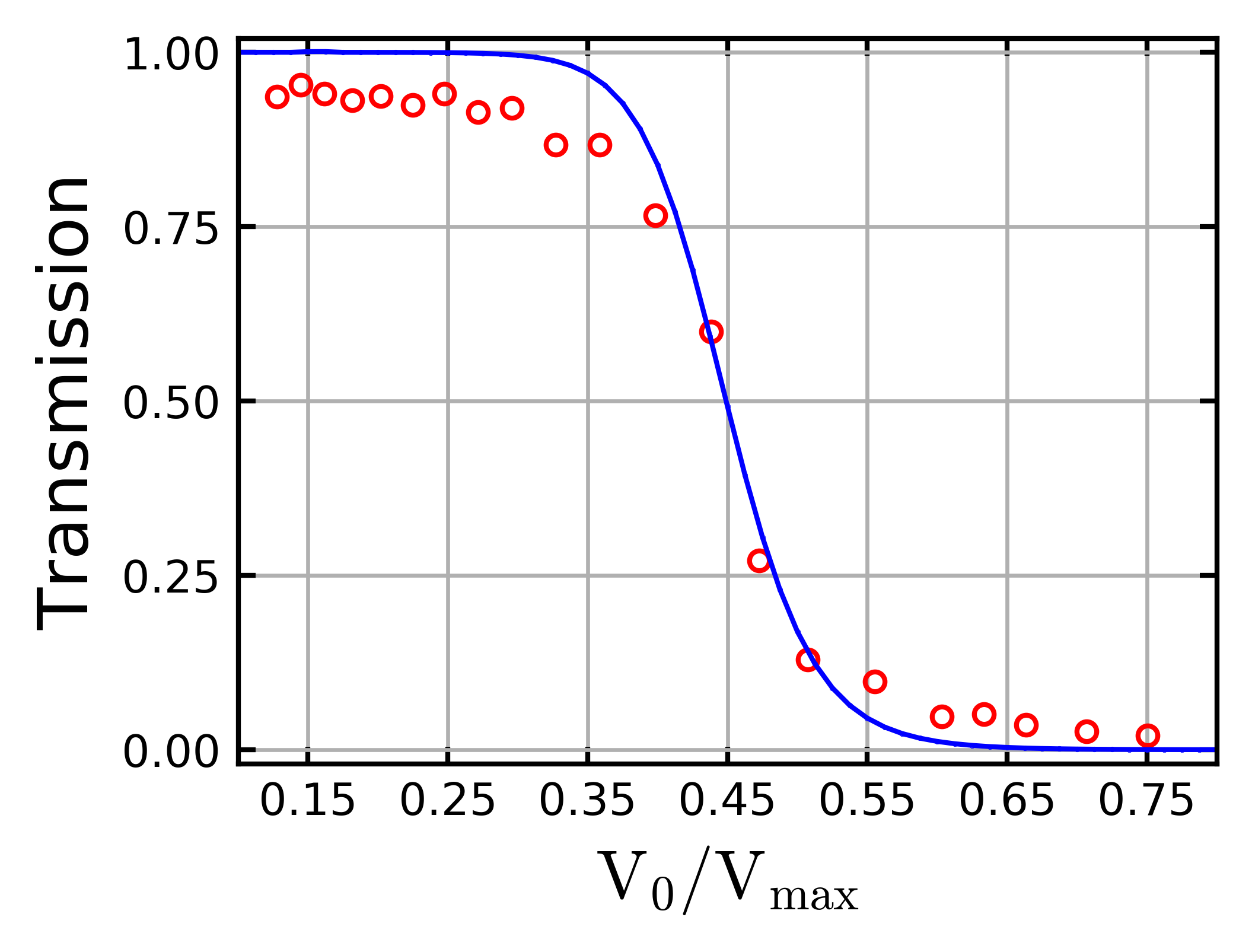}
\caption{Transmission in the classical regime $\sigma\sim 2\,\mu$m$>\lambda_{\text{dB}} = 1.2 \, \mu \text{m}$. The transmission is plotted as a function of the barrier height $V_0$ that is normalized to its maximum value $V_\textrm{max}$ obtained at full laser power. The points are the experimental data. The solid line is the curve expected from simulations, with the theoretical sinc PSF and assuming no velocity spread of the atoms. The barrier height is the only fit parameter. 
\label{Trans_func_V0/Vmax_10_pixels}}
\end{figure}

We first focus on a case of high atom velocity 8.6\,mm/s ($\lambda_\textrm{dB}=1.2\,\mu$m) and a barrier made of 10 pixels on the DMD. It corresponds to a size $\sigma_0=1.96\,\mu$m on the atoms. In that case, we expect diffraction and optical aberrations to only slightly enlarge the barrier width $\sigma \sim \sigma_0$. The transmission as a function of the barrier height is observed to be a step function although smoothed in the region of the transition (Fig.\,\ref{Trans_func_V0/Vmax_10_pixels}). Such a behavior is indeed expected in the classical regime, $\lambda_\textrm{dB}<\sigma$. The 50\% transmission is then obtained when the barrier height $V_0$ corresponding to the kinetic energy of the atoms. The data can be directly compared to numerical simulations that consist in solving the 1D time-independent Schr\"odinger equation for an incoming plane wave. The  ratio of the transmitted plane wave to the incoming plane wave amplitudes gives the transmission. The simulation globally matches the experimental data. The fact that the experimental transmission does not really go from 1 to 0 in our data is a consequence of thermal atoms whose behavior is not well captured in the simulation. For example, some thermal atoms due to their backward initial velocities have not encountered the barrier in the time of the experiment although they are counted as reflected. 

\begin{figure}[ht]
\centering
\includegraphics[width=0.9\columnwidth]{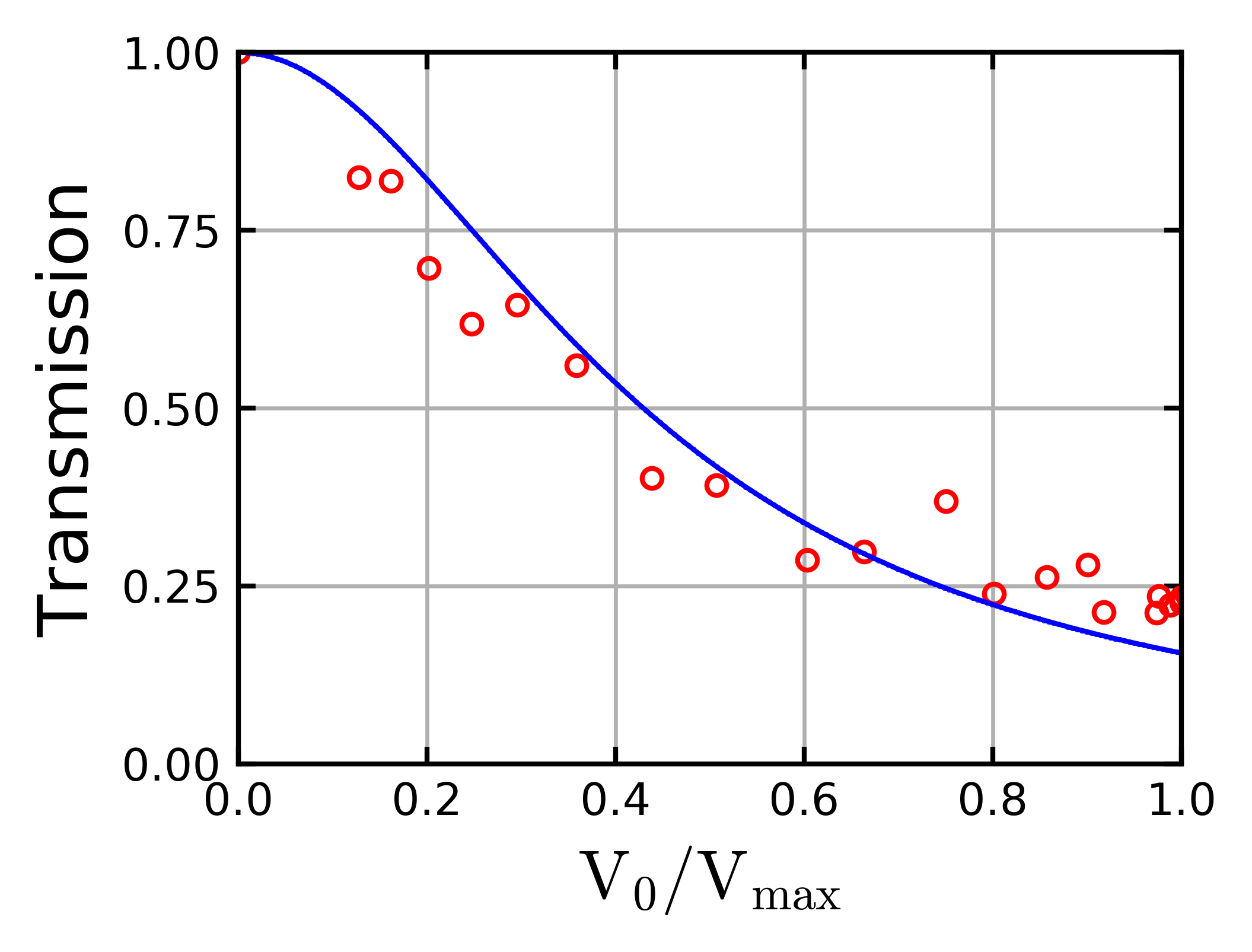}
\caption{Transmission in the quantum regime $\sigma\sim 1\,\mu$m$<\lambda_{\text{dB}} = 3.0\, \mu \text{m}$. The solid line corresponds to the expected transmission for a delta potential barrier $(1/(1+V_0^2/V_\textrm{ref}^2)$ and no velocity spread of the atoms. $V_\textrm{ref}$ is fitted to the data. 
\label{Trans_func_V0/Vmax_3_pixels}}
\end{figure}

The previous results can be compared to the situation at low atom velocity 3.7 mm/s ($\lambda_\textrm{dB}=3.0\,\mu$m) and a barrier made of 3 pixels on the DMD. The size of the potential barrier is then mostly given by the resolution of the imaging system. The transmission curve then qualitatively changes shape with a constantly decaying behavior. Such a behavior is expected in the quantum regime $\lambda_\textrm{dB}>\sigma$. In that case, the barrier width is not resolved by the atoms, the barrier can be theoretically replaced by a delta potential and the corresponding analytical transmission curve matches the experimental data. As previously, deviations from the theoretical transmission curve are probably coming from the thermal gas contribution (Fig.\,\ref{Trans_func_V0/Vmax_3_pixels}).

In both previous cases, the shape of the transmission curve is not very dependent on the actual PSF of the imaging system. Indeed, in the first case, the barrier width is dominated by the number of pixels creating the potential on the DMD, whereas in the quantum case, the width is not resolved by the atoms. In order to experimentally access the PSF and its width, we turn to a different experiment where we compare at a constant velocity (5\,mm/s), the measured 532\,nm power that is necessary to reflect 50\% of the atoms for different barrier widths (Fig.\,4). For our parameters (although we are not always strictly in the classical regime), the 50\% transmission is numerically found to well correspond the situations where the barrier height equals the kinetic energy $V_0=E_\textrm{kin}$. Such an experiment thus permits to measure the relative barrier maxima for barriers made of different numbers of pixels $N_\textrm{pix}$ or equivalently of different width $\sigma_0$. 
\begin{figure}[ht]
\centering
\includegraphics[width=\columnwidth]{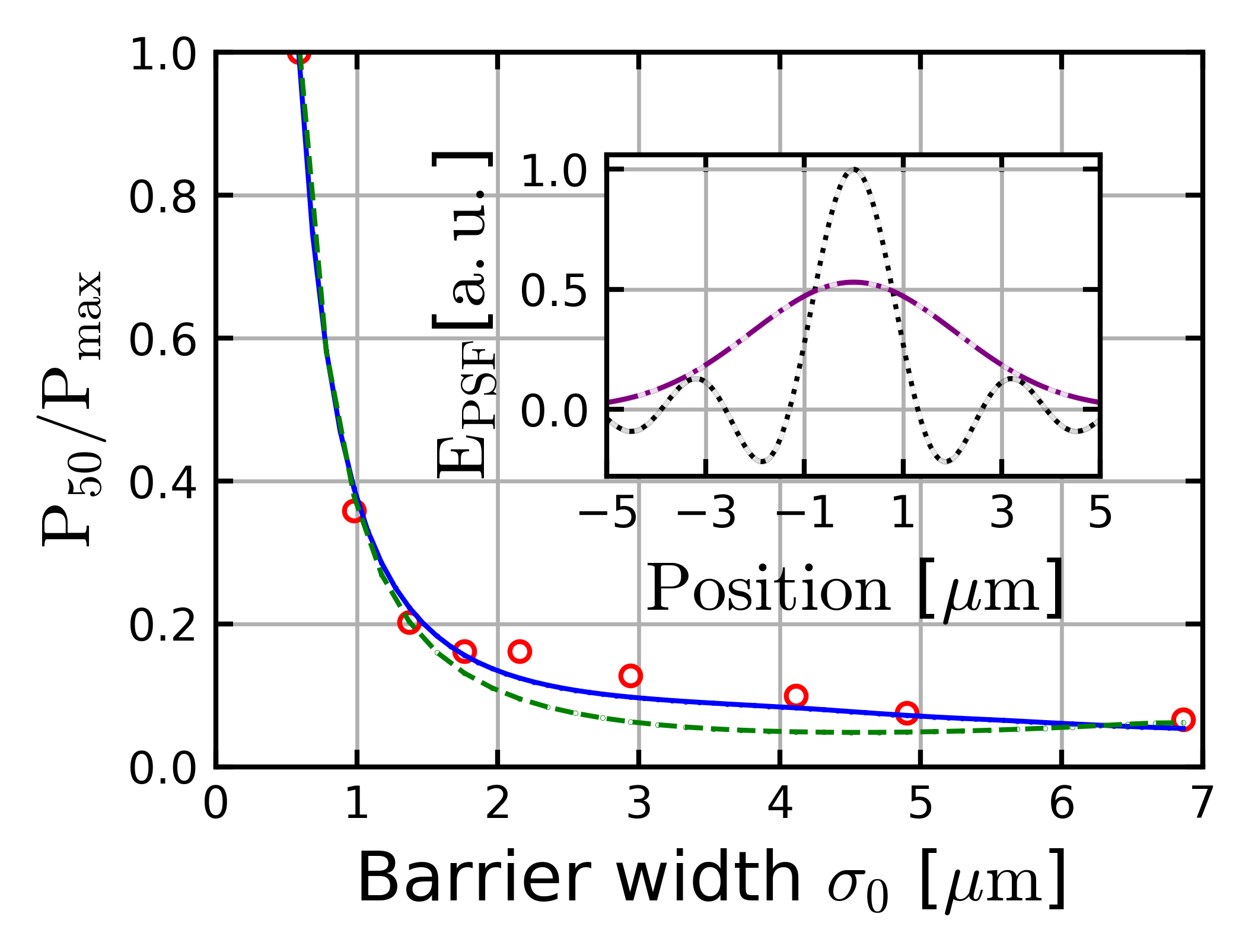}
\caption{Optical power $P_\textrm{50}$ needed to reflect 50\% of the atoms as a function of the barrier width $\sigma_0$. The optical power is normalized to its maximum value $P_\textrm{max}$. The mean atomic velocity is $5\,\text{mm/s}$, corresponding to a $\lambda_{\text{dB}} = 2.0 , \mu \text{m}$. The dashed (solid) line corresponds to the numerical result with a sinc PSF with $\sigma_\textrm{res}=2.0\,\mu$m (the ansatz PSF). The real (dotted-dashed line) and imaginary (dotted line) parts of the ansatz PSF are plotted in the inset (see text).  
\label{barrier_height_func_barrier_width_pixels}}
\end{figure}
The exact barrier shape is the absolute value square of the convolution of the amplitude PSF with door functions of variable width $\sigma_{0}=N_\textrm{pix} p$. In the limit of small $\sigma_0$, the barrier maxima is proportional to $\sigma_0^2$ as the field interferes constructively. In this case, the 50\% transmission power scales as $1/{\sigma_0^2}$. On the contrary, for large barrier size, we expect the barrier maximum height and thus also the 50\% transmission power to be independent of $\sigma_0$. These two limiting behaviors are experimentally observed (Fig. \ref{Trans_func_V0/Vmax_3_pixels}). The change of behavior occurs when the resolution of the imaging system is of the order of $\sigma_0$. Assuming a sinc PSF with a resolution $\sigma_\textrm{res}$, we can calculate the barrier maximum for each barrier width $\sigma_0$. We find that in order to reproduce the observed power ratio between the two limiting regimes, we need $\sigma_\textrm{res} \approx 2 \sigma_\textrm{th}=2.0\,\mu$m (dashed curve in figure \ref{barrier_height_func_barrier_width_pixels}). However, the whole curve is not well fitted, in particular for barrier widths between 2 and 5 microns. Our results thus point toward a PSF modified by optical aberrations, i.e. not a sinc function. The consideration of a more complex PSF as presented below permits to better match the observation (solid curve in figure 4).

\begin{figure}[ht]
\includegraphics[width=0.9\columnwidth]{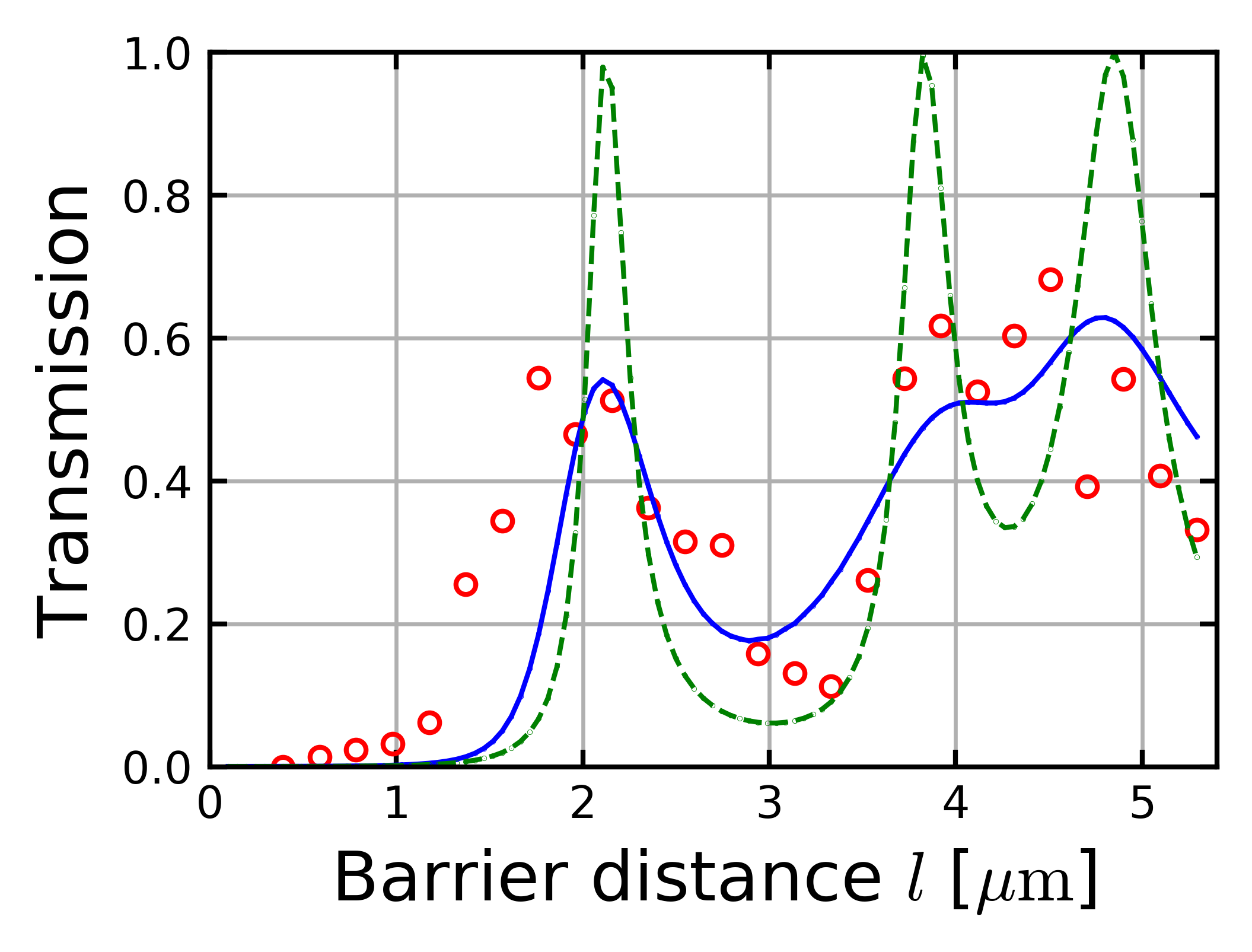}
\caption{Transmission through two barriers as a function of the distance between them. The mean atomic velocity is $\sim 3.3 \, \text{mm/s}$ ($\lambda_{\text{dB}} = 3.1\,\mu \text{m}$). The solid line is the numerical solutions with the ansatz PSF (see text) and a velocity spread $\Delta v=0.25\,$mm/s. The dashed line is the numerical solution for monokinetic atoms. In this last case, the Fabry-P\'erot resonance peaks are resolved.  
\label{Trans_func_pixels_2_pixels}}
\end{figure}

We now turn to transmission experiments through double barriers. In this series of experiments, we use a velocity of $3.3\,\text{mm/s}$, barrier widths of 2 pixels and 75\% of the maximum power available. For such parameters, the transmission through a single barrier is $\sim$70\%.  We then vary the distance $l$ between the two barriers and, as previously, study the transmission (Fig.\,\ref{Trans_func_pixels_2_pixels}). Interestingly, the transmission shows oscillations as a function of the barrier distance. However, one should be careful in directly interpreting them as Fabry-P\'erot peaks. First, in contradiction to the expectation, they do not reach 100$\%$ transmission. Second, the position the peaks are not precisely positioned at multiple of $\lambda_\textrm{dB}/2$. Note that for larger barrier distances, the transmission is essentially constant to $\sim 50\%$. 

In order to interpret our data, two important important ingredients have to be taken into account. First, the atom have a velocity spread that can wash out interference effects. It can be estimated to $\Delta v=0.25\,$mm/s from the experimental parameters and can be considered in a simulation by averaging the transmission over the velocity distribution. Second the studied barrier distances are not much larger than the resolution of the optical system. We should thus take into account that the barriers have some width and moreover that the optical fields originating from the two DMD regions may interfere as we use coherent light. In principle, the exact calculation of the optical profile requires the knowledge of the amplitude PSF of the optical objective (with its real and imaginary parts), which differs from the theoretical one. In preliminary measurements, prior to the installation on the atoms, the image of a point source (corresponding to the intensity PSF) appeared to be diffraction limited only up to a numerical aperture of $\sim 0.15$. The outer part of the objective only marginally contributed to the reduction of the size of the central image spot and was rather observed to add a broad background. We attribute this effect to imperfectly manufactured (non-spherical) lenses. A reasonable ansatz ampltitude PSF is thus a sinc function originating from the central part of the objective in addition to a Gaussian broader background originating from the outer part of the objective. For simplicity, we chose this Gaussian component to be imaginary such that there is not interference with the real sinc function.  

With such an ansatz, we can calculate the optical potential and simulate the experimental transmission. We then adjust the PSF parameters to reproduce the experimental findings of both figure 4 and 5. We find that the amplitude PSF
$\textrm{E}_\textrm{PSF}\propto \text{sinc}(\pi x/\sigma_\textrm{sinc})+ iA_\textrm{G}\exp(-x^2/2\sigma_\textrm{G}^2)$ with $\sigma_\textrm{res}=1.30\,\mu$m, $\sigma_\textrm{G}=2.07\,\mu$m, and $A_\textrm{G}=0.53$ (see inset in Fig.\,4) permits to reasonably match the data (Fig.\,\ref{barrier_height_func_barrier_width_pixels} and \ref{Trans_func_pixels_2_pixels}). Although, the actual PSF certainly differ from our ansatz, we expect the latter to capture its main features. Using the previous ansatz PSF, we also plot in figure 5, the expected transmission for monokinetic atoms. In that case, the transmission does reach $100\%$ for specific barrier distance corresponding to Fabry-P\'erot resonances. The precise positions of the resonance depend on the atom velocity leading to a broadening of the transmission peaks after velocity averaging. We thus find that the first Fabry-P\'erot resonance peak is resolved but not the second and third ones which merge into a single broad peak. Note also that the monokinetic resonance peaks are not equally spaced. This is due to optical interference between the optical fields originating from the different areas on the DMD. The precise potential barrier heights and positions are slightly affected by the positive to negative oscillations of the amplitude PSF. In particular, for a barrier distance of $\sim 1.9\,\mu$m ($\sim 3.2\,\mu$m), the barrier heights are minimum (maximum), because the sinc function is maximally negative (positive) at those distances. Overall, we thus attribute the observed oscillations in the transmission to a combination of Fabry-P\'erot interference and optical interference due to an oscillatory PSF. 

In conclusion, we have realized some textbook experiments on atomic transmission through optical barriers. Key advantages of our experiments are the use of non-interacting Bose-Einstein condensates and of a digital micro-mirror device for the tuning of the barrier characteristics. We have clearly demonstrated both the classical and quantum regimes of atomic scattering on a single optical barrier by adjusting both the barrier width and the De Broglie wavelength, or equivalently, the speed of the atoms. We then have realized an atomic Fabry-P\'erot interferometer based on two potential barriers and have been able to resolve the first Fabry-P\'erot resonance peak. Interestingly, our setup using a DMD to produce the optical barriers allow for experiments with varying barrier widths and distances permitting an $in-situ$ estimation of the PSF of the optical system. Technical improvements, such as further reduction of the atom velocity spread using delta-kick cooling technique \cite{Amman97} and/or the engineering of moving barriers using the high switching rate capabilities of DMDs \cite{Ha_2015, Gauthier_2021} could lead to the observation of a complete atom Fabry-P\'erot resonator spectrum. 

In the future, the capacity to realize coherent beam splitters from single barrier could be important in the context of atom interferometry. In addition, it open prospects for the study of the influence of interactions on collisions with potential barriers \cite{Martin_2007,Manju_2018, Lindberg23}. Scattering of a bright soliton on a barrier have been predicted to lead to non-classical NOON states \cite{Weiss09, Cornish_2009, Boisse17, Naldesi23}. In the case of a Fabry-P\'erot interferometer, interactions are expected to lead to quantum effects such as squeezing of the outcoming atomic cloud or atomic blockade \cite{Carusotto01}.

%\begin{acknowledgements}
This  research  has  been  supported  by  CNRS,  Minist\`ere  de  l'Enseignement  Sup\'erieur  et  de  la  Recherche, Labex PALM, Quantum Paris-Sacaly, R\'egion Ile-de-France  in  the  framework  of  Domaine d'Int\'er\^et Majeur Quantip, Paris-Saclay in the framework of IQUPS, ANR Droplets (19-CE30-0003), Simons foundation (award number 563916:  localization of waves).
%\end{acknowledgements}

%\bibliographystyle{IEEEtran} 
\bibliography{mybibliography} 
\end{document}